# Drivers of Mobile Payment Acceptance: The Impact of Network Externalities in Nigeria

**Qasim M. Ajao[1*], Olukotun Oludamilare[1], Lanre Sadeeq[2]**

[1]Faculty of Electrical and Computer Engineering, Georgia Southern University, Statesboro, GA, USA
[2]Microsoft Corporation, Redmond, USA
Email: *Qasim.ajao@ieee.org

## Abstract

The rising popularity of mobile payments can be attributed to the widespread use of smartphones and their applications. Despite its potential to simplify our lives, its adoption in African countries has been limited. This paper aims to enhance our understanding of the critical factors that influence the acceptance of mobile payments in Nigeria by exploring the impact of "network externalities" in addition to conventional technology acceptance factors. It posits that performance expectancy, effort expectancy, social influence, trust, and network externality are the key drivers of mobile payment acceptance. The research findings indicate that while traditional drivers still have an impact on customers' willingness to adopt mobile payment, network externalities have the strongest influence. The paper provides recommendations for future research, although the results did not support the impact of effort expectancy.

## Subject Areas

Information and Communication, Security, Privacy, and Trust

## 1. Introduction

Mobile payment refers to the financial service that allows individuals to carry out transactions using their mobile devices. This service has gained popularity in recent years, especially due to the emergence of smartphones and mobile applications. While some argue that mobile payment is a mere means of accessing internet payment services through mobile devices, others observe that the context differs, despite the similarity in functionality. In Nigeria, internet usage is high, with over 70% of the population having internet access [1]. Additionally, many Nigerians own smartphones, creating a solid foundation for mobile payment adoption. Nonetheless, the adoption rate of mobile payment in Nigeria has been relatively slow compared to other countries such as China, the United States, and some European countries. Although small mobile payments are becoming more prevalent, widespread adoption has been impeded by concerns regarding security, standardization, and inconsistent user experience [2]. Nonetheless, experts forecast that the mobile payment market will keep expanding in Nigeria, and it is expected to reach $ 465 billion in 2030 [3].

Compared to the United States and China, Nigeria, being the "giant of Africa", currently lacks the infrastructure and consumer base required to support the growth of mobile payment adoption. Moreover, further enhancements in security, standardization, and user experience are necessary to increase the widespread usage of mobile payments. There is a need for additional research on mobile payment acceptance, particularly in the Nigerian context where the technology is accessible, but adoption is still limited. This paper employs the Unified Theory of Acceptance and Use of Technology (UTAUT) to explore the factors that influence the intention to use mobile payment. UTAUT includes four primary predictors: performance expectancy, effort expectancy, social influence, and facilitating conditions [4].

However, since technology adoption models are specific to the technology being studied, this paper also incorporates two additional constructs. Trust is a significant factor in financial transactions, especially when conducted over wireless networks, and it has been shown to affect the intention to use technology. Additionally, this paper considers network externality, which is particularly relevant in developing countries where users' choice of mobile network can impact their use of mobile technology applications. This paper's structure is as follows: the literature review section examines UTAUT, trust, and network externalities, as well as mobile technology and mobile payment in a fourth subsection. The third section describes the empirical model, also known as the methodology, followed by data analysis and further discussion in section four. The paper concludes with recommendations and conclusions for future research.

The primary objective of investigating technology adoption is to acquire a thorough comprehension of the factors that impact individuals' choices to adopt a particular technology. The Unified Theory of Acceptance and Use of Technology (UTAUT), developed by Venkatesh *et al.* [5], is a well-known model in this field. However, when analyzing the adoption of mobile payment technology, it is critical to consider the influence of factors such as trust and network externalities. In the subsequent three sections, we will explore the three elements of the proposed conceptual framework that we plan to adopt in detail.

## 2. UTAUT, Network Externalities, and Trust

The Unified Theory of Acceptance and Use of Technology (UTAUT) was developed by Venkatesh *et al.* [5], and it integrates eight theories and models, including TAM, MM, MPCU, TPB, TRA, IDT, SCT, and C-TAM-TPB. UTAUT is considered a comprehensive model for technology acceptance, and it includes dynamic influences through the addition of four moderating variables, which enhance its explanatory abilities. The model consists of four main constructs, namely performance expectancy, effort expectancy, social influences, and facilitating conditions [6] [7] [8]. UTAUT is a suitable model for cross-cultural studies, as it is sensitive to cultural aspects, and it has been extended to include hedonic motivation, price value, and habit. Empirical investigations have shown that UTAUT outperforms competing models in studying the factors influencing technology acceptance and behavioural intentions [5] [6].

Effort expectancy, performance expectancy, facilitating conditions, and social influences are identified as important predictors of technology acceptance. Effort expectancy refers to the ease of use and complexity of a technology, which indirectly impacts behavioural intentions through performance expectancy. Performance expectancy includes perceived usefulness, relative advantage, and outcome expectation, and it is considered one of the strongest predictors of technology acceptance. Facilitating conditions refer to the technical and organizational infrastructure that makes using technology easy, and social influences refer to the pressure exerted by an individual's social surroundings to perform or not perform a behaviour [9] [10].

Network externalities are the phenomenon where the perceived value of a product or service increases as the number of users increases. Direct, positive, and indirect network externalities are the three types of network externalities that explain the relationship between utility and the number of buyers. Trust is a critical factor in the acceptance of mobile payment technology, particularly in developing countries [11] [12]. Trust refers to the willingness of one party to be vulnerable to the second party and allow it to conduct important actions on its behalf. Inexperienced customers tend to build their perceptions and opinions of other people in their social surroundings [13]. Trust is an under-investigated variable in the context of mobile payment technology, and it is essential for the technology's acceptance. Boosting customer trust can significantly predict mobile payment adoption when combined with other factors such as perceived usefulness, perceived ease of use, price, and peers' influence [14] [15].

### Previous Studies—Mobile Payment

The UTAUT model has been extensively used to investigate the factors that contribute to the adoption of mobile payments [4]. Effort expectancy, performance expectancy, social influence, and facilitating conditions have been identified as factors that directly impact customers' mobile payment intentions, according to previous studies. Most of the studies that employed the UTAUT model used survey research as the primary method of data collection [16].

Dahlberg and Mallat conducted a qualitative study to explore the impact of network externalities on mobile payment adoption [17]. They found that network externalities play a role in this process, but the external validity of their findings was limited due to a small sample size. Subsequently, Mallat used focus groups to further investigate the impact of network externalities on mobile payment adoption, but the generalization of the results was still an issue [18] [19]. The term "mobile payment" generally refers to payments made using mobile devices with phone capabilities, although it can include all mobile devices. For this study, any form of activity initiation, activation, and confirmation is considered a form of mobile payment [20].

Mobile payments are divided into two categories: proximity payments and remote payments, depending on the customer's location, relationship with the merchant, and usage scenario. Proximity payments involve exchanging credentials within a small distance using RFID technology or bar-code scanning and are also known as point-of-sale payments. Among proximity payment technologies, NFC is the most promising due to its convenience and security. NFC devices offer three operating modes: peer-to-peer mode, reader/writer mode, and card emulation mode, enabling contactless payments or ticketing. Remote payments are similar to online shopping and are conducted through mobile web browsers or smartphone applications [4] [21] [22]. Although remote payments seem to be more mature than proximity payments, both types can be integrated to improve the mobile payment market in the future.

## 3. Empirical Approach and Model

This study chose the UTAUT model due to its comprehensive approach to technology acceptance theories and its high explanatory power [5].

The Unified Theory of Acceptance and Use of Technology (UTAUT) is a theoretical model developed by Venkatesh *et al.* to understand individuals' acceptance and usage of technology. It integrates four key constructs: performance expectancy, effort expectancy, social influence, and facilitating conditions. Performance expectancy relates to how individuals believe technology can enhance their job performance, while effort expectancy refers to the perceived ease of using the technology. Social influence encompasses the influence of peers and colleagues, and facilitating conditions involve the availability of resources and support for technology adoption. These constructs interact to shape individuals' behavioral intentions and actual usage of technology. The UTAUT model has been widely employed in various fields to gain insights into technology acceptance and usage patterns, assisting researchers and practitioners in designing and implementing technology that is more likely to be embraced and utilized. However, it is important to consider the model's applicability may vary across different technologies and contexts, prompting researchers to tailor it to their specific research objectives and settings [5].

The goal of this research is to examine the factors that impact customers' intentions to use mobile payments. Through a literature review, certain constructs were identified as essential predictors of mobile payment acceptance that were not initially included in the UTAUT model. Conversely, some constructs in the model were discovered to have no significant influence on customers' acceptance of mobile technology. To improve the UTAUT model's explanatory power, trust was integrated, as it is a crucial factor in financial transactions literature. As with many mobile services, network externalities play a role in mobile payment adoption [20]. As this study is focused on behavioural intention, the facilitating conditions construct, one of the original UTAUT constructs, was removed from the model as it was not deemed to have a significant impact on mobile payment acceptance. Figure 1 outlines the proposed research model.

### 3.1. Empirical Objective and Research Hypotheses

The objective of this paper is to investigate the factors influencing the intention to use mobile payment methods. To achieve this, a literature review was conducted to identify the significant factors affecting behavioural intentions (BI). The proposed set of hypotheses based on the research model is discussed below. Effort expectancy refers to the perceived ease of use associated with a particular technology [5]. When an individual perceives using a technology as easy, they are more likely to use it [24]. Prior studies have demonstrated that customers' belief that using mobile payment methods is effortless strengthens their intentions to use them [4].

Therefore, the following hypothesis is proposed:

H1: Effort expectancy **(EE)** has a positive effect on customers' intentions
**(BI)** to use mobile payment methods.                                    (1)

The concept of Performance Expectancy refers to an individual's perception of the degree to which a technology can enhance their performance [5]. If a user

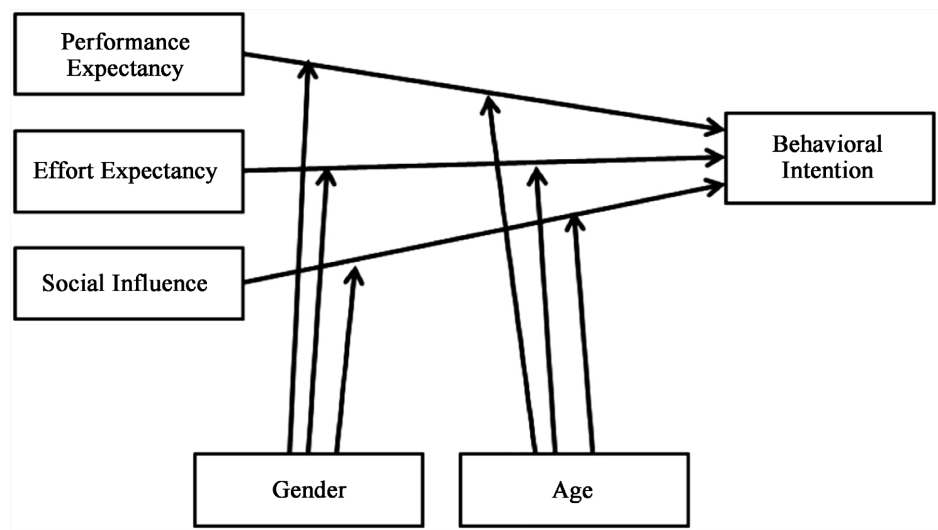

**Figure 1.** Empirical model.

perceives that a technology will improve their performance, they are more likely to use it [25]. The existing literature on mobile payment acceptance suggests that users are more likely to adopt mobile payment if they believe it will be helpful [4]. Therefore, the following hypothesis is proposed:

**H2**: Performance Expectancy **(PE)** positively influences customers' intentions **(BI)** to use mobile payment. (2)

Social Influence is the pressure exerted on an individual by their social network to use or not to use an innovation [5]. This factor has been shown to be a significant predictor of technology acceptance in various contexts. In the area of mobile payment acceptance, research confirms that social pressure affects a user's willingness to use mobile payment systems [4]. Thus, the following hypothesis is proposed:

**H3**: Social Influence **(SI)** positively influences customers' intentions **(BI)** to use mobile payment. (3)

Trust in the online environment refers to a customer's willingness to be vulnerable to a vendor after evaluating the vendor's characteristics, who is expected to provide an agreed-upon service [26]. Trust influences a customer's likelihood of accepting a given technology [27]. In the mobile payment context, customers' trust in the security and reliability of mobile payment systems is expected to drive their acceptance of it [28] [29]. Thus, the following hypothesis is proposed:

**H4**: Trust positively influences customers' intentions **(BI)** to use mobile payment. (4)

Network externalities occur when the benefits of using a product increase as more people use it [30]. The network externality effect, as identified by Van Hove, is a significant factor affecting payment systems. In the context of mobile payments, customers' adoption decisions are influenced by the number of users of a technology [31], with a large customer base considered a necessary condition for adoption. Moreover, the availability of mobile payment services at a greater number of merchants also increases customers' willingness to adopt them [17]. Additionally, the more merchants that offer mobile payment services, the more willing customers are to adopt them [32]. Thus, the following hypothesis is proposed:

**H5**: The number of mobile payment users (merchants and customers) positively influences customers' intentions to use mobile payment. (5)

The model depicted in Figure 1 was utilized to employ the UTAUT model to investigate the acceptance of mobile payment technology. This includes the impact of age and gender on the four fundamental factors of the UTAUT model, including effort expectancy, performance expectancy, facilitating conditions, and social influence, and found that all four factors significantly influenced the acceptance of mobile payment technology in the specific region. This shows the importance of considering demographic factors such as age and gender when evaluating the acceptance and usage of new technologies across different countries [33].

## 3.2. Model Development

This study utilized a questionnaire to measure the proposed constructs in the developed model. To ensure the questionnaire's validity, the items were adapted from existing literature and customized to fit the context of mobile payment technology. The questionnaire utilized a 5-point Likert scale ranging from strongly disagrees to strongly agree. As the original questionnaire was in the Arabic language (refer to **Appendix Table A1**), it was necessary to translate it into English for use in Nigeria. To ensure the translation's accuracy, back translation was employed [34]. Furthermore, to guarantee that the translated questions were culturally appropriate and easily understood by the specific population being studied, the questionnaire underwent back translation [35]. The final translated questionnaire was then distributed to the survey participants.

## 3.3. Survey and Data Collection

Recruiting participants for a study in Nigeria can be challenging and is influenced by various factors such as research design and target population. Similarly, this study adopted a meticulous approach to recruit participants and distribute questionnaires to ensure a representative sample that accurately reflects the research objectives. Recruitment methods used include personal distribution, contact persons, or online survey platforms depending on specific factors.

As in previous studies, participation was voluntary, and incentives were used to encourage participation. The target sample size was 300, with most surveys administered in person. To aid in data collection, six contact individuals consisting of secondary school teachers, university undergraduates, graduate students, and researchers themselves were utilized. The research team randomly selected participants in natural settings such as schools and assessed their familiarity with mobile payments before administering the survey. The survey was distributed between July 4th and 15th, 2022, with no incentives given to participants. Although the target sample size was not achieved, 258 out of 280 surveys were retrieved, and 253 were considered usable after a preliminary visual assessment.

Despite some deviation from distribution percentages, the study participants were considered representative of the Nigerian population, including individuals from various age groups, educational levels, and average income levels. Ensuring a representative sample and the validity and reliability of survey instruments are essential in producing trustworthy and generalizable research findings, despite any differences between previous studies and research in Nigeria.

### Implication of Independent Variables

1) Effort expectancy is a crucial variable in UTAT as it influences the likelihood of technology adoption and use. Designers and developers should prioritize making the technology easy to use and minimizing the perceived effort required to encourage adoption.

2) Social influence plays a significant role in UTAT, as it can positively or negatively affect technology adoption and use. Incorporating social elements into the technology can encourage positive social influence and increase the likelihood of adoption by leveraging the perception of others using and benefiting from the technology.

3) Performance expectancy is another important variable in UTAT, as individuals need to believe that using the technology will lead to improved performance. Designers and developers should focus on demonstrating the benefits and value of the technology to enhance performance expectancy and increase the likelihood of adoption.

4) Trust is a critical factor in technology adoption and use. Building trust in the technology by ensuring its security, reliability, and providing transparent information is essential to encourage individuals to adopt and use the technology.

5) Network externalities, which stem from the benefits derived from the number of people using the technology, are influential in UTAT. Increasing the network externalities of the technology can enhance its value and subsequently drive adoption and use. Promoting the technology through various channels, such as social media, can increase its visibility and attract more users, thereby increasing network externalities.

## 4. Analysis and Discussion on Data Collection

After collecting 253 surveys, the data was entered into an Excel sheet and imported into SPSS and AMOS 20 for analysis to answer the research questions and test the hypotheses. The upcoming sections will explain the data analysis and discuss the obtained results.

For this study, the data collected was entered into an Excel spreadsheet and subsequently imported into software tools such as SPSS and AMOS 20 for analysis. These software packages were used to conduct the necessary statistical analyses and test the research questions and hypotheses formulated for the study. By utilizing these analytical tools, researchers were able to analyze the data and draw conclusions based on the results obtained. SPSS is widely used for statistical analysis, while AMOS is specifically designed for structural equation modeling, allowing researchers to examine complex relationships among variables. The use of these software tools facilitated the rigorous analysis of the data, enhancing the reliability and validity of the study findings.

### 4.1. Initial Analysis

To assess the potential impact of outlier data on the results, the study conducted an initial multiple regression test on 253 responses, after excluding two cases based on case-wise diagnostics Table 1 displays the demographic characteristics of the sample, including information on gender, age, education, and income.

The sample comprises more females, middle-aged individuals, bachelor's degree holders, and middle-income subjects. The second preliminary analysis examined the correlations between independent variables, which can affect the magnitude and direction of betas in the regression equation.

The analysis involved two tests, inspection of the correlation matrix and collinearity statistics. To test for collinearity, the study computed tolerance and variance inflation factor (VIF), as presented in Table 2. Tolerance values measure how much a variable influences other independent variable, while VIF values measure how much the variance of each regression coefficient increases when independent variables are correlated. The study concluded that there were no issues with multicollinearity, as all tolerance and VIF values in Table 2 were within acceptable limits. The internal consistency between the survey items measuring each independent variable and the dependent variable was assessed using Cronbach's *α*. In social sciences, alpha values greater than 0.6 are considered acceptable for reliability. Table 3 reports the reliability measures, which indicate that the survey items used to measure each variable were highly consistent.

Table 1. Collected data sample.

| Gender | Number | % | Age | Number | % |
|---|---|---|---|---|---|
| Men | 67 | 30.7% | Below 20 | 59 | 23.5% |
| Women | 180 | 67.7% | Between 20 – 40 | 146 | 58.2% |
| Data not captured | 4 | 1.6% | Above 40 | 39 | 15.5% |
| Total | 251 | 100% | Data not captured | 7 | 2.8% |
| | | | Total | 251 | 100% |
| Education | Count | % | | | |
| Secondary School | 37 | 10.8% | Income *(in Nigerian Naira)* | Number | % |
| B.Sc. Holder | 150 | 63.7% | Below N250,000.00 | 46 | 22.3% |
| Graduate Degree | 25 | 12% | N250,000.00 - N550,000.00 | 119 | 47.4% |
| Other Certificates | 34 | 9.5% | Above N550,000.00 | 66 | 22.3% |
| Data not captured | 5 | 4% | N/A | 20 | 8% |
| Total | 251 | 100% | Total | 251 | 100% |

Table 2. Collinearity statistics.

| Independent variable | Tolerance | VIF |
|---|---|---|
| Effort expectancy | 0.533 | 1.877 |
| Social influence | 0.698 | 1.432 |
| Performance expectancy | 0.513 | 1.950 |
| Trust | 0.667 | 1.500 |
| Network externalities | 0.550 | 1.819 |

Table 3. Cronbach's *α* (including factoring the means and standard deviations).

| Variable Data | Item | Mean | Standard deviation | Cronbach's *α* | Number |
|---|---|---|---|---|---|
| Effort expectancy | Q-1 | 3.54 | 1.17 | 0.857 | 4 |
| | Q-2 | 3.55 | 0.96 | | |
| | Q-3 | 3.59 | 1.02 | | |
| | Q-4 | 3.68 | 1.15 | | |
| Performance expectancy | Q-5 | 3.51 | 1.42 | 0.805 | 4 |
| | Q-6 | 3.44 | 0.79 | | |
| | Q-7 | 3.03 | 0.25 | | |
| | Q-8 | 3.40 | 1.22 | | |
| Social influence | Q-9 | 3.06 | 0.79 | 0.750 | 3 |
| | Q-10 | 3.01 | 0.44 | | |
| | Q-11 | 3.40 | 1.74 | | |
| Trust | Q-12 | 2.76 | 1.07 | 0.723 | 4 |
| | Q-13 | 3.02 | 1.55 | | |
| | Q-15 | 3.22 | 1.25 | | |
| Network externalities | Q-16 | 3.92 | 1.11 | 0.740 | 3 |
| | Q-17 | 3.23 | 0.78 | | |
| | Q-18 | 3.55 | 1.11 | | |
| Behavioral intentions | Q-19 | 3.14 | 1.54 | 0.876 | 3 |
| | Q-20 | 3.38 | 0.78 | | |
| | Q-21 | 3.28 | 1.20 | | |

## 4.2. VIF Analysis

A confirmatory factor analysis was conducted to comprehend how the variables measure the factors they signify. The outcomes of this analysis, presented in Table 4, include the factor loadings, construct reliability, average variance extracted, maximum shared variance, and average shared variance. Factor loadings exceeding 0.5 are deemed significant, and except for Q14, all factors exhibit significant loadings.

The construct reliability (CR) and average variance extracted (AVE) measures suggest that all the model constructs possess reliable and convergent features. Model constructs with CR values above 0.7 are deemed reliable, and all the model constructs meet this standard. Furthermore, constructs with an AVE greater than 0.5 indicate that the construct's variance is more significant than the variance caused by error, and all the model constructs satisfy this requirement. Additionally, the results reveal that the factors within each construct are not excessively correlated, indicating a good level of convergent validity. Discriminant validity occurs when the factors measuring each construct have more correlations with each other than with other factors. To establish discriminant validity, the average shared variance (ASE) and maximum shared variance (MSV) must be lower than the AVE. In this study, the measurement model fulfils these criteria.

**Table 4.** Result analysis of the factors.

| Data Information | | Loaded Factor | #Reliability Of Construct | Extracted Variance-#Average | Variance-Shared #Maximum | Variance-Shared #Average |
|---|---|---|---|---|---|---|
| Effort Expectancy | Q-1 | 0.74 | 0.872 | 0.643 | 0.442 | 0.282 |
| | Q-2 | 0.82 | | | | |
| | Q-3 | 0.80 | | | | |
| | Q-4 | 0.81 | | | | |
| Performance Expectancy | Q-5 | 0.70 | 0.828 | 0.556 | 0.542 | 0.328 |
| | Q-6 | 0.80 | | | | |
| | Q-7 | 0.62 | | | | |
| | Q-8 | 0.66 | | | | |
| Social Influence | Q-9 | 0.75 | 0.750 | 0.506 | 0.265 | 0.222 |
| | Q-10 | 0.63 | | | | |
| | Q-11 | 0.74 | | | | |
| Trust[1] | Q-12 | 0.90 | 0.804 | 0.611 | 0.337 | 0.215 |
| | Q-13 | 0.62 | | | | |
| | Q-15 | 0.73 | | | | |
| Network Externalities | Q-16 | 0.83 | 0.802 | 0.585 | 0.452 | 0.316 |
| | Q-17 | 0.76 | | | | |
| | Q-18 | 0.62 | | | | |
| Behavioral Intentions | Q-19 | 0.87 | 0.858 | 0.644 | 0.442 | 0.348 |
| | Q-20 | 0.82 | | | | |
| | Q-21 | 0.73 | | | | |

[1]Item #14 has been deleted because of the result analysis.

### 4.3. Evaluating the Model Fit Measure and Testing of the Hypotheses

Before testing the hypotheses, it is crucial to examine the rationale behind the variables. Table 5 presents the correlation matrix showing the significant relationships between the independent variables, namely performance expectancy, effort expectancy, social influence, trust, and network externalities, and the de-pendent variable, behavioural intentions. Performance expectancy and network externalities have the most considerable bivariate impact on behavioral intentions among these variables. All the correlation values in Table 5 are acceptable, indicating a significant correlation between behavioral intentions and each independent variable.

To assess the suitability of the research model, we utilized AMOS 20 and evaluated the values of chi-square, degree of freedom, CFI, and RMSEA, which are shown in Table 6 along with their acceptable thresholds. All reported values fall within the acceptable range, indicating a good fit for the research model. The model is capable of predicting behavioral intentions, with the exception of effort expectancy, and has an explanatory power of 58%. Table 7 shows the standardized

**Table 5.** Matrix data-correlation.

| The Constructs Data | EE | PE | SI | T | NE | BI |
|---|---|---|---|---|---|---|
| Effort Expectancy denoted: (EE) | 1 | | | | | |
| Performance Expectancy denoted: (PE) | 0.637** | 1 | | | | |
| Social Influence denoted: (SI) | 0.376** | 0.408** | 1 | | | |
| Trust denoted: (T) | 0.362** | 0.418** | 0.476** | 1 | | |
| Network Externalities denoted: (NE) | 0.588** | 0.571** | 0.365** | 0.440** | 1 | |
| Behavioral Intentions denoted: (BI) | 0.545** | 0.615** | 0.469** | 0.544** | 0.632** | 1 |

**At a level of 0.01 denoted as (2-tailed) the Correlation is significant.

**Table 6.** Model fit-value evaluation.

| Index | Value | Threshold Values (Source: Hair *et al.* 2009) |
|---|---|---|
| Chi-square | 333.82, $P < 0.001$ | Value is Significant |
| Degree of freedom | 149 | - |
| $X^2$/df (deg. of freedom) | 2.215 | <0.302 |
| CFI | 0.916 | >0.809 |
| TLI | 0.917 | >0.808 |
| RSMEA | 0.058 | <0.06 remain significant if the sample size is >250 and the CFI > 0.902 |

**Table 7.** The relationships sets and corresponding standardized betas.

| Hyp# | Relationship | Result | Stand. Beta |
|---|---|---|---|
| H1 | Effort expectancy → Behavioral intentions | Not Supported | 0.04 |
| H2 | Performance expectancy → Behavioral intentions | Supported | 0.35 |
| H3 | Social Influence → Behavioral intentions | Supported | 0.13 |
| H4 | Trust → Behavioral intentions | Supported | 0.22 |
| H5 | Network externalities → Behavioral intentions | Supported | 0.34 |

beta coefficients for the estimated relationships, as well as the results of hypothesis testing. The prediction equation is provided below:

$$BI = 0.35\,PE + 0.13\,SI + 0.22\,T + 0.34\,NE + error \tag{6}$$

### 4.4. Discussion of Results

The results of the research model were largely positive, with the exception of effort expectancy (H1), which did not have a significant impact on customers' behavioral intentions towards mobile payment, despite a significant correlation in the bivariate relationship (as indicated in Table 5). This implies that while ease of use is somewhat important to Nigerian customers, its importance is diminished by the high penetration and daily use of mobile phones for this relatively new mobile-based technology. This finding is consistent with previous studies but differs from some other research [36] [37].

In contrast, customers' acceptance of mobile payment was significantly predicted by both performance expectancy (H2) and social influence (H3). The findings suggest that Nigerian customers value the potential benefits of mobile payment and are influenced by the opinions of others in their social circle, which is consistent with prior studies. Trust (H4) and network externalities (H5), two other variables included in the research model, were also found to be significant predictors of mobile payment acceptance. For customers to conduct financial transactions via mobile payment, they must have confidence in the technology and the service provider, which is where trust (H4) comes in. Network externalities (H5) had the most significant impact on customers' behavioral intentions towards mobile payment, accounting for 23.7% of the variance [37].

This result indicates that customers are more likely to use mobile payment when more merchants accept this payment method, and they believe that the more people using it, the lower the cost. As previous research has shown, creating a critical mass is essential for driving customers' acceptance of mobile payment [17]. Table 7 displays the estimated relationships and their corresponding standardized beta, along with the results of hypothesis testing. The prediction equation is presented below, and the findings are summarized in the table.

### 5. Conclusion and Future Development

In this study, the influence of five major predictors on customers' intentions to use mobile payment technology was investigated: performance expectancy, effort expectancy, social influence, trust, and network externality. The bivariate correlations confirmed a significant relationship between each predictor and behavioral intentions. However, when all predictors were considered together in the research model, some predictors lost their significance due to commonalities that may be attributed to an unknown factor. The results revealed that all predictors, except for effort expectancy, significantly predicted behavioral intentions, providing support for all five hypotheses. Notably, network externality had the strongest impact on mobile payment acceptance. Together, the five predic-tors explained 58% of the variance in behavioural intentions.

## 5.1. Implications and Recommendations

This paper discusses several implications for future research in the field of mobile payment acceptance. The following recommendations should be considered:

- The influence of effort expectancy (ease of use) on mobile payment acceptance should be studied further through path analysis, as it did not have a direct influence in the model but may have an indirect influence through performance expectancy.
- Network externalities should be included as a major predictor in future mobile payment acceptance studies, as it is an important factor for technology acceptance but not commonly included in technology acceptance models.
- Performance expectancy remains an essential predictor of technology acceptance and should be included in any model, as it is a dominant construct in technology acceptance research. Merchants, banks, and other businesses should monitor the factors influencing the adoption of mobile payment technology and design their marketing policies accordingly. Trust is also a key factor for customers' willingness to accept mobile payment services, and managing the organization's image and creating a trustworthy brand name should precede offering mobile payment services.
- System developers should prioritize performance expectancy when designing mobile payment systems, as it is the second most influential predictor of acceptance. To improve uptake, developers should aim to maximize the number of payment types supported, the ability to handle different currencies, and processing speed.

## 5.2. Limitations

Governments across Africa have been promoting the use of mobile payments to combat the spread of the pandemic, with some even waiving transaction fees. Kenya, where mobile money was first developed in Africa, has been particularly successful in this regard. According to the Economic and Financial Affairs Council, Kenya recorded a record-high of $55.1 billion in mobile transactions in 2021, which is almost a 20% increase from the previous year. The Global Consumer Survey conducted by Statista also revealed that 84% of Kenyan internet users used mobile phones for payments, which is higher than in Europe. Although only a quarter of the Kenyan population has internet access, there is still room for growth in mobile payments. In Nigeria, where internet penetration is around 34%, 60% of internet users made mobile payments in 2021 [38] [39] [40].

In African countries like Kenya and Nigeria, mobile payments play a crucial role in promoting financial inclusion, especially for those without access to traditional banking services. Safaricom's M-Pesa, launched in Kenya in 2007, is the leading mobile wallet in Africa, with over 50 million active monthly users.

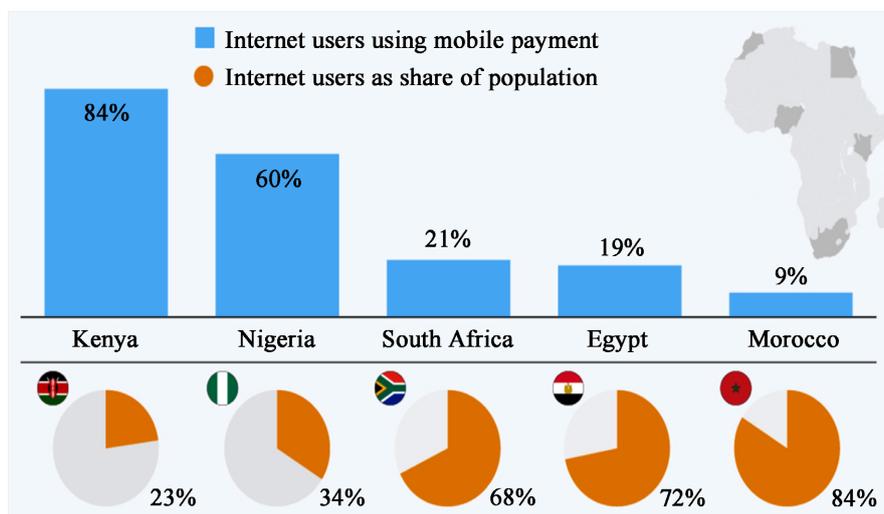

**Figure 2.** Potential of Mobile Payment in Africa as at 2021.

Statista's Global Consumer Survey found that 84% of Kenyan internet users made mobile payments in 2021, while in Nigeria, 60% of internet users utilized mobile payments, despite low internet penetration rates (**Figure 2**). The ease of use and accessibility of mobile payments, as well as reduced transaction times, are some of the benefits they offer. However, challenges such as inadequate infrastructure and a need for trustworthy agents can hinder their adoption. Researchers have identified factors such as relative advantage, perceived ease of use, compatibility, trust, and security as significant determinants of intention to use mobile payments. Consumers who are already comfortable with internet and mobile banking are more likely to use mobile payments and encourage others to do so. A secure and reliable service is essential for the successful transition to a cashless society in Nigeria [4].

## Conflicts of Interest

The authors declare no conflicts of interest.

# Appendix

**Table A1.** Original Sources for items used to build the instrument.

| Item | Source |
| --- | --- |
| *Effort expectancy* | |
| √ Learning how to use mobile payment is easy for me. | Venkatesh *et al.* (2012); Peng *et al.* (2011) |
| √ My interaction with mobile payment is clear and understandable. | |
| √ I find mobile payment easy to use. | |
| √ It is easy for me to become skillful at using mobile payment. | |
| *Performance expectancy* | |
| √ I find mobile payment useful in my daily life. | Venkatesh *et al.* (2012); Peng *et al.* (2011) |
| √ Using mobile payment increases my chances of achieving things that are important to me. | |
| √ Using mobile payment helps me accomplish things more quickly. | |
| √ Using mobile payment increases my productivity. | |
| *Social influence* | |
| √ People who are important to me think that I should use mobile payment | Venkatesh *et al.* (2012); Peng *et al.* (2011) |
| √ People who influence my behavior think that I should use mobile payment. | |
| √ People whose opinions that I value prefer that I use mobile payment. | |
| *Trust* | |
| √ Mobile payments are trustworthy | Zmijewska *et al.* (2004) |
| √ I believe that data sent is confidential. | |
| √ I get an immediate confirmation message of the transaction. | |
| √ I trust mobile payment systems to be reliable. | Gefen (2000); Jarvenpaa *et al.* (2003) |
| *Network externalities* | |
| √ *If more and more merchants accept mobile payment, then:* | |
| √ The quality of mobile payment services will improve. | Yu and Tao (2007); Katz and Shapiro (1992) |
| √ A wider variety of mobile payment services will be offered. | |
| √ Customers will have to pay less to use mobile payment services. | |
| *Behavioral intentions* | Venkatesh *et al.* (2012); Peng *et al.* (2011) |
| √ I intend to use mobile payment in the future. | |
| √ I expect that I will use mobile payment in my daily life. | |
| √ I plan to use mobile payment frequently. | |